%% file: manuscript.tex
\documentclass[%
preprint,
 amsmath,amssymb,
aps,
]{revtex4-2}
\usepackage{float}
\usepackage{graphicx}
\usepackage{dcolumn}
\usepackage{bm}
\usepackage{subfigure}
\usepackage{diagbox}

\begin{document}

\preprint{APS/123-QED}

\title{Topological properties of two-dimensional photonic square lattice without $C_4$ and $M_{x(y)}$ symmetries}

\author{Langlang Xiong$^{1,3}$}

\author{Yufu Liu$^{2,3}$}%

\author{Yu Zhang$^{2,3}$}%

\author{Yaoxian Zheng$^{4}$}%

\author{Xunya Jiang$^{1,2,3}$}%
 \email{jiangxunya@fudan.edu.cn}
 \affiliation{
	$^{1}$Institute of Future Lighting, Academy for engineering and technology, Fudan University, Shanghai, 200433, China
}%
\affiliation{
	$^{2}$Department of Illuminating Engineering and Light Sources, School of Information Science and Engineering, Fudan University, Shanghai, 200433, China
}%
\affiliation{
	$^{3}$Engineering Research Center of Advanced Lighting Technology, Fudan University, Ministry of Education, Shanghai, 200433, China
}%
\affiliation{
	$^{4}$THz Technical Research Center and College of Physics and Optoelectronic Engineering, Shenzhen University, Shenzhen, 518060, China
}%

\date{\today}

\begin{abstract}
Rich topological phenomena, edge states and two types of corner states, are unveiled in a two-dimensional square-lattice dielectric photonic crystal without both $C_4$ and $M_{x(y)}$ symmetries. Specifically, non-trivial type-\uppercase\expandafter{\romannumeral1} corner states, which do not exist in systems with $C_4$ and $M_{x(y)}$ since the degeneracy, are protected by non-zero quadrupole moment, no longer quantized to but less than $0.5$. Excellent properties, e.g. sub-wavelength localization and air-concentrated field distribution, are presented. Type-\uppercase\expandafter{\romannumeral2} corner states, induced by long-range interactions, are easier realized due to asymmetry. This work broadens the topological physics for the symmetries-broken systems and provides potential applications.

\end{abstract}

\maketitle


\section{Introduction}

Topological quantum and classic systems with extraordinary properties, such as robust edge states, have been a growing interest \cite{lu2014topological,bansil2016colloquium, shun2018topological, ozawa2019topological, wang2020topological,kim2020recent}, which are the driving force of technology innovation\cite{wu2017applications}. Recently, a new class of topological systems, namely higher-order topological insulators (HOTIs) \cite{frank2018higher, zhang2019second, zhang2020low, xie2021higher, liu2021bulk}, has been demonstrated by theories and experiments. The most focused for HOTIs is the zero-dimensional (0D) corner state of two- or three-dimensional systems. In the research of topological systems, symmetries play a critical role generally \cite{chiu2016classification}. For example, non-zero Berry curvatures generally exist in a system without time-reversal symmetry or spatial inversion symmetry, namely Chern insulators\cite{wang2008reflection, wang2020universal} or valley insulators\cite{ma2016all, kim2021multiband,xi2020topological}, which admit the emergence of edge states.
Moreover, a two-dimensional (2D) HOTI with square lattice, the topological invariant $Q_{c}$, also known as quadrupole moment or fractional corner charge, is quantized to 0 or 0.5 when a system presents mirror symmetries $M_x:=x \to -x$ and $M_y := y \to -y$ or fourfold rotation $C_4$ symmetry, leading to topological non-trivial corner states that are protected by non-zero $Q_c$ \cite{benalcazar2017electric,benalcazar2017quantized,wheeler2019many}.

As for high order topology, the corner states can be divided into two types. Type-\uppercase\expandafter{\romannumeral1} corner states are protected by non-zero multiple moments. In particular, HOTIs with vanishing dipole densities but non-zero quadrupole moments, namely quadrupole topological insulators (QTIs), because of certain crystalline symmetries, e.g., reflection and rotation \cite{benalcazar2017quantized, benalcazar2019quantization}, the non-zero quadrupole moment is quantized to 0.5, which promises to the existence of type-\uppercase\expandafter{\romannumeral1} corner states. Very recently, the photonic QTIs have been realized by a square lattice without time-reversal symmetry \cite{he2020quadrupole}. 
Type-\uppercase\expandafter{\romannumeral2} corner states are caused by the long-range coupling between two edge states. For instance, a photonic kagome lattice with non-trivial 2D Zak phase\cite{liu2018topological, liu2017novel}, and by increasing the long-range interactions, i.e., coupling beyond next-nearest-neighbor(NNN) hopping, two sets of type-\uppercase\expandafter{\romannumeral2} corner states with different spatial symmetry can distinguish from the spectrum of edge states\cite{ni2019observation, li2020higher, shen2021investigation, wang2021higher, xu2020general}. However, in contrast to HOTIs in the systems with ``perfect lattice symmetries", e.g. the $C_3$ symmetry for kagome lattices and the $C_4$ symmetry for square lattices, the phenomena of HOTIs with breaking of the perfect lattice symmetries have not been intensively investigated to the best of our knowledge. The reason for lacking such research maybe because the topological invariant, multipole moment, can not be quantized in these systems, so that they are widely regarded as topologically trivial.
It would be very novel to reveal that, even with the breaking of perfect lattice symmetries, such as the photonic square lattice without both $C_4$ and $M_{x(y)}$ symmetries, rich phenomena of HOTIs and the topological edge states still can be observed. Furthermore, if we gradually make the system more and more asymmetric from the original system with perfect lattice symmetries, the continuous evolving of HOTIs and the topological edge states in this process are also very inspiring, since the topological origin of HOTIs, the varying trend of $Q_c$  and the Berry curvature from non-zero to zero can be carefully investigated.

In this work, we systemically investigate the topological properties of 2D photonic square lattice without both $C_4$ and $M_{x(y)}$ symmetries. We first construct a photonic square lattice with perfect $C_4$ lattice symmetry since four rods in one cell are identical.
Then by gradually changing the dielectric constant of two diagonal rods, the $C_4$ and $M_{x(y)}$ symmetry are broken and the topological phase transition appears, such as the annihilation of topology-degenerate singularities that carry non-zero Berry curvatures, leading to two sets of edge states.
Surprisingly, two types of corner states appear in the process without the perfect lattice symmetries.
Because of the asymmetries of lattice, type-\uppercase\expandafter{\romannumeral1} corner states which are protected by non-zero quadrupole moments, is no longer quantized to but less than 0.5, and type-\uppercase\expandafter{\romannumeral2} corner states that caused by long-range interactions can distinguish from the above edge states easily. We also find that a whole gap range can only exist type-\uppercase\expandafter{\romannumeral1} corner states but no edge states, and larger long-range interactions could cause added type-\uppercase\expandafter{\romannumeral2} corner states. The above results from the strict numerical methods are also confirmed by the results from the tight binding model, which gives more clear physical origins of the two types of corner states.
This work will prove valuable in expanding the understanding of topological phases beyond ``perfect lattice symmetries". Furthermore, these findings could be friendly to the application of edge states and corner states due to the all-dielectric structure.

\section{Edge states}

\begin{figure}[htbp]
	\centering
		\includegraphics[width=0.45\textwidth]{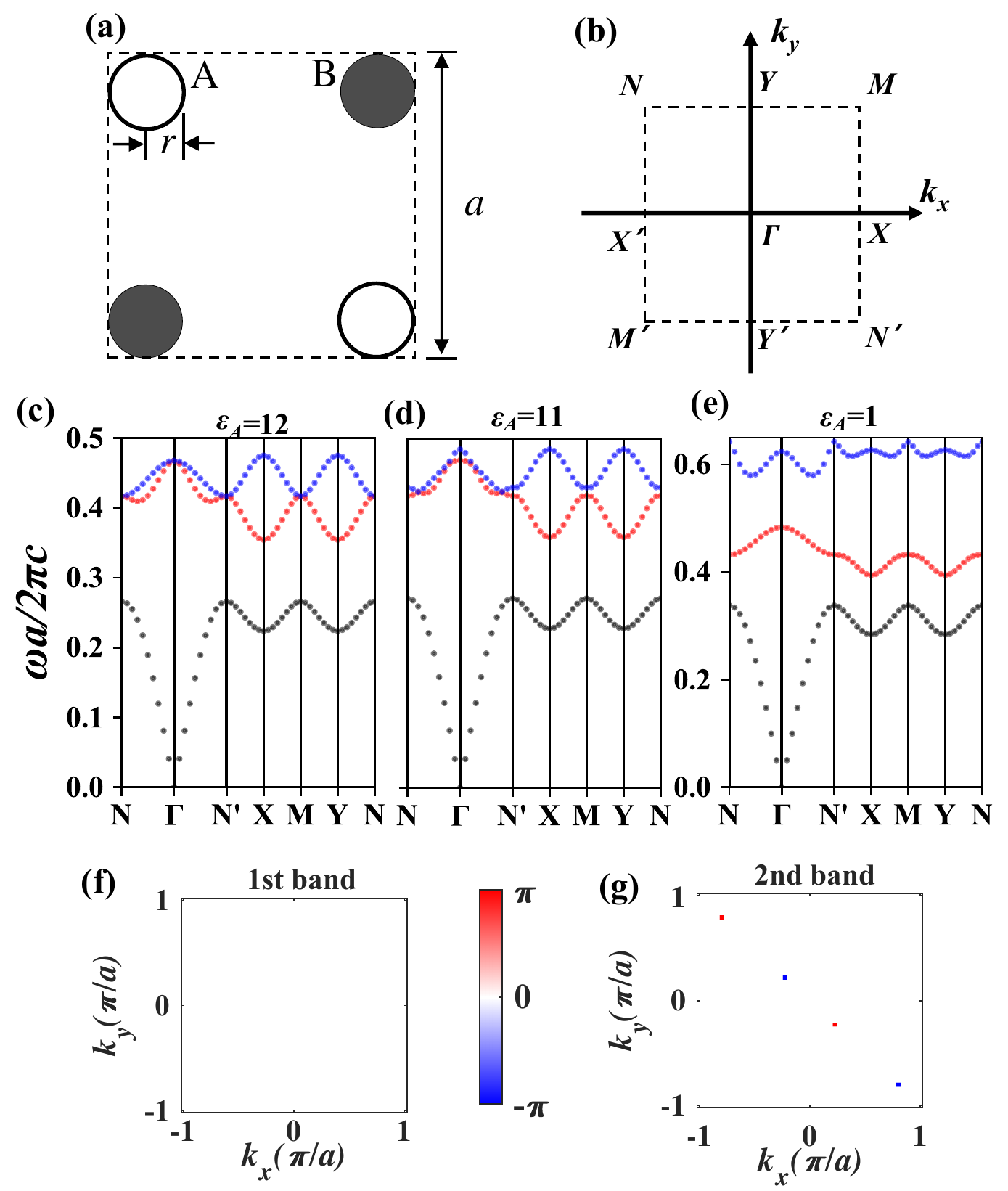}
	\caption{(a) A schema of 2D photonic unit cell with square lattice, the radius of all-dielectric rods is fixed as $r=0.12a$, where $a$ is the lattice constant; (b) The first Brillouin zone of square lattice; (c-e) Band structures of PhCs with parameter $\varepsilon_A=12$ (c), $11$ (d), and $1$ (e), while $\varepsilon_B$ is fixed as 12. (f-g) Distributions of Berry curvatures for the first and second band with parameters $\varepsilon_A=11$, $\varepsilon_B=12$, $r=0.12a$.}\label{model}
\end{figure}

We start by considering the 2D photonic unit cell with square lattice as shown in Fig. \ref{model}(a).
The four corners of the unit cell contain four dielectric rods in the air with radius $r=0.12a$, where $a$ is the lattice constant. The top left and down right rods are marked as type-\textit{A} with the relative permittivity $\varepsilon_A$ and permeability $\mu_A$, while the others are marked as type-\textit{B} rods with $\varepsilon_B$ and $\mu_B$. The first Brillouin zone of this PhC is illustrated in Fig. \ref{model}(b). At the first step, we suppose $\varepsilon_A=\varepsilon_B=12$, $\mu_A = \mu_B = 1$, and Fig. \ref{model}(c) shows the band structure of transverse electric (TE, $Ez$ polarization) mode. Because of spatial symmetry and time-reversal symmetry, the second and third bands degenerate at $\Gamma$ and $M(N)$ points.

Our goal is to investigate the topological phase transition when the square lattice is without both $C_4$ and $M_{x(y)}$ symmetries. Hence, we turn $\varepsilon_A$ of type-\textit{A} rods slightly from 12 to 11, and the band structure is shown in Fig. \ref{model}(d). Since $C_4$ and $M_{x(y)}$ symmetries are broken, the degeneracies at $\Gamma$ and $M(N)$ points are opened, and those degenerate points, namely topological singularities, move along $\Gamma - N$ direction.

To research the topological phase of the second gap, we calculate the Berry curvatures which are normalized to $[-\pi, \pi]$ for the disturbed model, and the details of the calculations are given in supplementary information S1. The Berry curvatures of the first and second bands are shown in Fig. \ref{model}(f) and Fig. \ref{model}(g). We find that two topological singularities with opposite non-zero Berry curvatures generate from $\Gamma$ point, while two other opposite topological singularities generate from $N$ point in Fig. \ref{model}(g). The sum of Berry curvatures, i.e., Chern number, keeps zero because of time-reversal symmetry. However, the sum of local Berry curvatures, namely valley Chern number, is non-zero, and we mark it as general valley Chern number $C_{gv}^{(n)}$ because there are degenerate points in concerned bands. We can calculate the general Chern number of the second band near $\Gamma$ point $C_{\Gamma}^{(2)}=\pm 0.5$ and near $N$ point $C_{N}^{(2)}=\pm 0.5$. The general valley Chern number of a single gap can be calculated by summing up $C_{gv}^{(n)}$ of all band(s) below the gap, which means the second gap has non-trivial $C_{gv}^{(2)}=\pm 1$. Because of the antisymmetry of the distributions for Berry curvatures as shown in Fig. \ref{model}(g), we can assume that the difference for the valley Chern number of the second gap between this model and its counterpart, i.e., $\varepsilon_A=12, \varepsilon_B=1$, is $\Delta C_{gv}^{(2)}=\pm 2$, which means two different topological edge states are supported according to the bulk-edge correspondence.

\begin{figure}[htbp]
	\centering
		\includegraphics[width=0.45\textwidth]{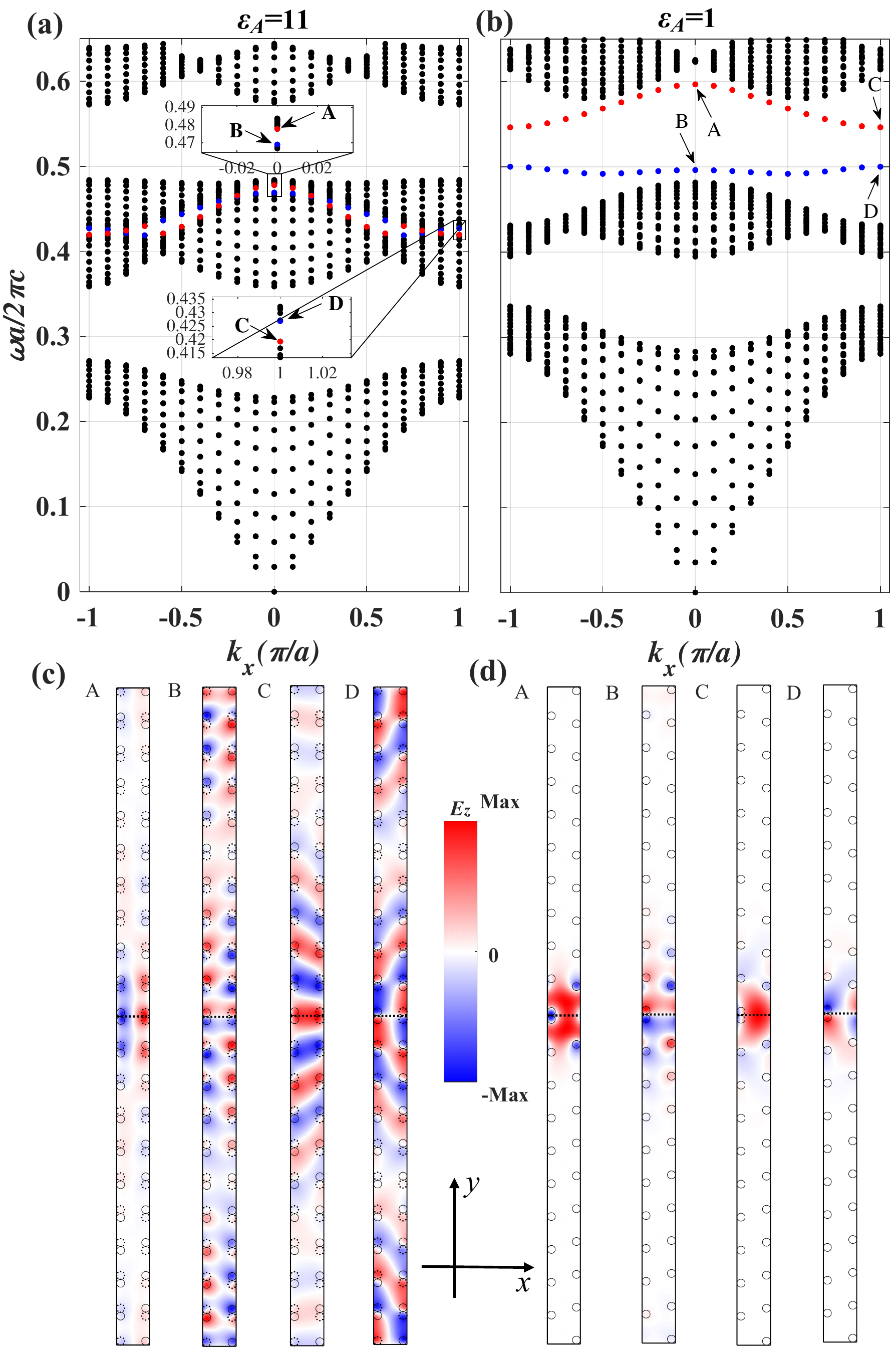}
	\caption{(a) Band structures of a supercell, the bulk states, symmetrical edge states, antisymmetrical edge states are marked as black, red, and blue dots, respectively. The parameters are set as above(below) 10 unit cells with $\varepsilon_{A(B)}=11$ and $\varepsilon_{B(A)}=12$. (b) The same as (a) but $\varepsilon_{A(B)}=1$. (c) Spatial distributions of $E_Z$ field for $k_x=0$ (A-B) and $k_x=\pi/a$ (C-D) with $\varepsilon_{A(B)}=11$, the circles of dash line mean dielectric rods with $\varepsilon_{A(B)}=11$, and circles of full line mean dielectric rods with $\varepsilon_{B(A)}=12$. (d) The same as (c) but $\varepsilon_{A(B)}=1$, the circles of full line mean dielectric rods with $\varepsilon_{B(A)}=12$}\label{EdgeState}
\end{figure}

To confirm the bulk-edge correspondence, we combine 10 unit cells whose parameters are $\varepsilon_A=11$, $\varepsilon_B=12$ above the boundary with its counterpart $\varepsilon_A=12$, $\varepsilon_B=11$ below the boundary. By using finite-element method(FEM) software COMSOL Multiphysics and the wave optics module, the band structure of a supercell with Bloch boundary condition along the x-direction and periodic boundary condition of continuity along the y-direction is shown in Fig. \ref{EdgeState}(a). Two sets of edge states in the second gap are marked as red dots with symmetry modes and blue dots with antisymmetry modes, and the $E_z$ fields with $k_x=0$ and $k_x=\pi /a$ are drawn in Fig. \ref{EdgeState}(c). However, the second gap with these parameters is not omnidirectional, meaning that the edge states would be hard to observe in the laboratory.

Fortunately, the width of the second gap will keep broadening if we decrease $\varepsilon_A$ continuously and fix $\varepsilon_B=12$. For example, when $\varepsilon_A=1$ and $\varepsilon_B=12$, the omnidirectional second gap is highly obvious as shown in Fig. \ref{model}(e). In the process of decreasing $\varepsilon_A$ from $11$ to $1$, some interesting phenomena are observed, such as two topological singularities with non-zero Berry curvature generate from $\Gamma$ point and two generate from $N$ point consolidate and annihilate by each other, so the Berry curvatures of second band change from non-zero to zero. According to the bulk-edge correspondence, the non-zero Berry curvatures admit the existence of edge states. Surprisingly, even Berry curvatures change to zero because of the annihilation of singularities, the two sets of edge states with spatial symmetry and antisymmetry alway exist in the process, which do evolve from the non-trivial edge states and are also protected by the chiral symmetry \cite{orazbayev2018chiral}, e.g., Fig. \ref{EdgeState}(b) shows the band spectrum of a supercell contains upper PhCs with $\varepsilon_A=1$, $\varepsilon_B=12$ and lower PhCs with $\varepsilon_A=12$, $\varepsilon_B=1$. The above results for the evolution of Berry curvatures and edge states with $\varepsilon_A$ changing are shown in supplementary information S2.

It would be interesting to analyze the two sets of edge states from the perspectives of physic and engineering.
First, the band spectrum of antisymmetrical edge states is almost flat, which can be demonstrated by two views: one is the strong antisymmetrical fields which concentrate on dielectric rods as mode B and D in Fig. \ref{EdgeState}(d) show, the other is the limited intra-cell interaction (see supplementary information S5 for details). The flat edge band can be used to produce the topological slow light \cite{arregui2021quantifying} and high-Q cavity. We mark these edge states as dielectric-edge-states (DESs). Second, the symmetrical edge states whose energies mainly concentrate on air as mode A and C in Fig. \ref{EdgeState}(d) show, which would be valuable in the topological waveguide to reduce the effect of the impurities in the medium and the light with high energies. Such edge states are marked as air-edge-states (AESs). More importantly, the coupling between the two sets of edge states can realize the topological Fano resonator \cite{zangeneh2019topo}.

\section{Corner states and fractional charge}

\begin{figure*}[htbp]
	\centering
		\includegraphics[width=0.9\textwidth]{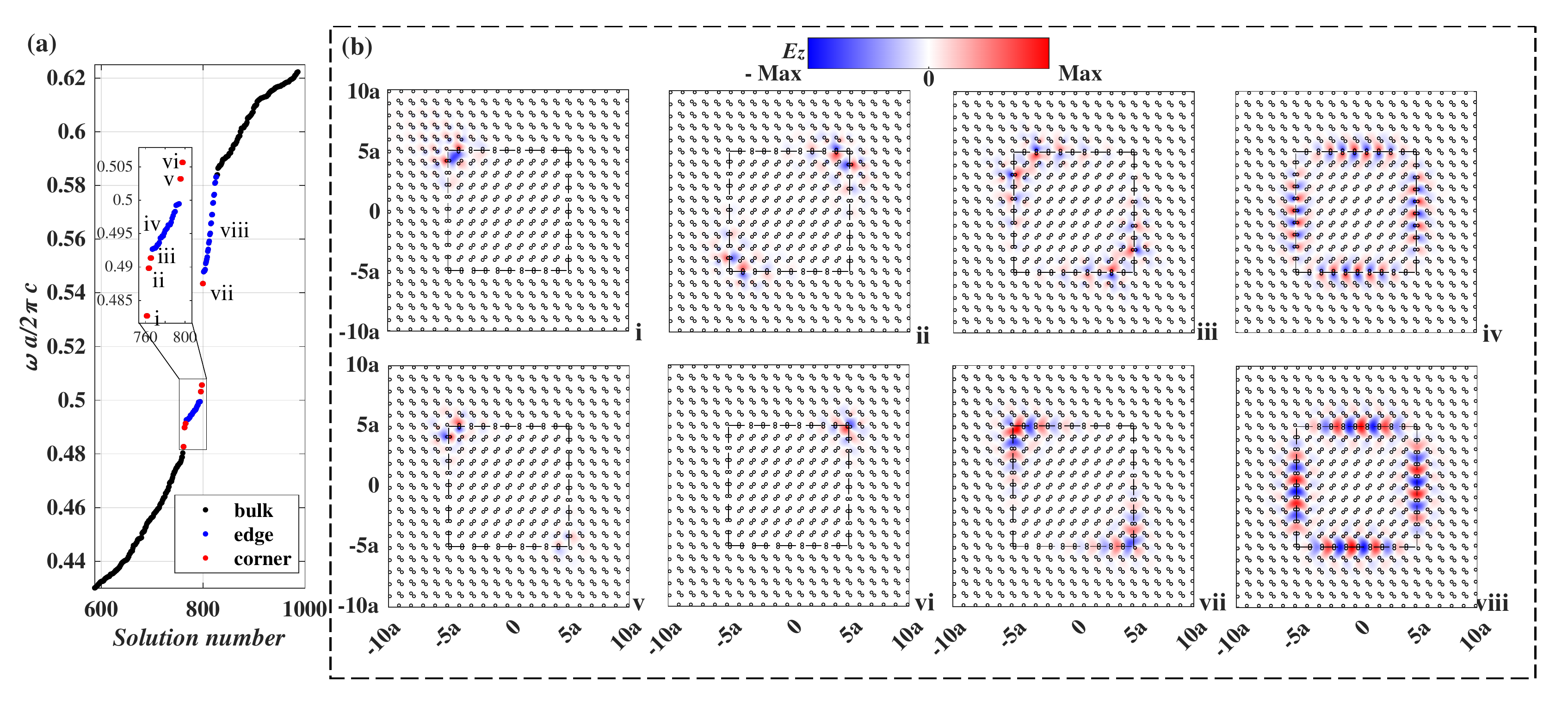}
	\caption{(a) Band structure of a combined supercell, the bulk, edge, and corner states are marked as black, blue, and red dots, respectively. Parameters are set as: the inside $10 \times 10$ unit cells with $\varepsilon_A=1$, $\varepsilon_B=12$ and the outside 5 layers of unit cells with $\varepsilon_A=12$, $\varepsilon_B=1$. (b) The spatial distributions for $E_z$ field of corner states(i-iii and v-vii) and edge states (iv and viii)}\label{CornerState}
\end{figure*}

In this section, we will focus on HOTI with corner states, namely 0D localized modes in the 2D system. The corner states can be divided into two types: type-\uppercase\expandafter{\romannumeral1} corner states from the eigenmodes of the unit cell, namely the ``zero-energy" mode of tight binding model, and type-\uppercase\expandafter{\romannumeral2} corner states caused by the long-range interactions of edge states.
So far two types of corner states have not been realized simultaneously in a PhC with $C_4$ symmetry and time-reversal symmetry, because of the degeneracies of bands \cite{SI} and the limited NNN hopping. However, if we construct a PhC without both $C_4$ and $M_{x(y)}$ symmetries, can we realize high order topological states and novel topological phases?  Next, we will demonstrate that interesting phenomena and special topology in such symmetries-broken systems, which is beyond the systems with perfect lattice symmetry, can be found.

At the first step, we combine two types of PhCs to investigate the corner states and edge states. As shown in Fig. \ref{CornerState}, a supercell of $10\times 10$ unit cells with $\varepsilon_A=1$ and $\varepsilon_B=12$ is surrounded by $5$ layers of unit cells with $\varepsilon_A=12$ and $\varepsilon_B=1$, and absorbing boundary condition is used in the outsides to avoid redundant states in the concerning gap. The spectrum around the 2nd gap is obtained by FEM and shown in Fig. \ref{CornerState}(a), in which bulk, edge, and corner states are marked by black, blue, and red dots, respectively. For convenience, we select eight typical states from low frequency to high frequency which are marked as i-viii, and the $E_z$ field distribution of those states are also shown in Fig. \ref{CornerState}(b). We start from two distinctive edge states DES-iv and AES-viii shown in Fig. \ref{CornerState}(b)-iv and -viii respectively as we mentioned in Section 2: DES-iv is antisymmetrical along the boundary of the two types of supercell and its field concentrates on the dielectric rods, while AES-viii is symmetrical and its field concentrates on the air. More specifically, the corner states near DES-iv in Fig. \ref{CornerState}(a) have similar symmetry and distribution features, such as states i, ii, iii, v, and vi, whereas the corner states near AES-viii have similar features, such as state vii.
From the view of long range coupling between different edges, the states i, ii, iii, and v can be judged as typical type-\uppercase\expandafter{\romannumeral2} corner states. The states vi and vii are type-\uppercase\expandafter{\romannumeral1} corner states which will be clearly demonstrated in the next model.

\begin{figure*}[htbp]
	\centering
		\includegraphics[width=1\textwidth]{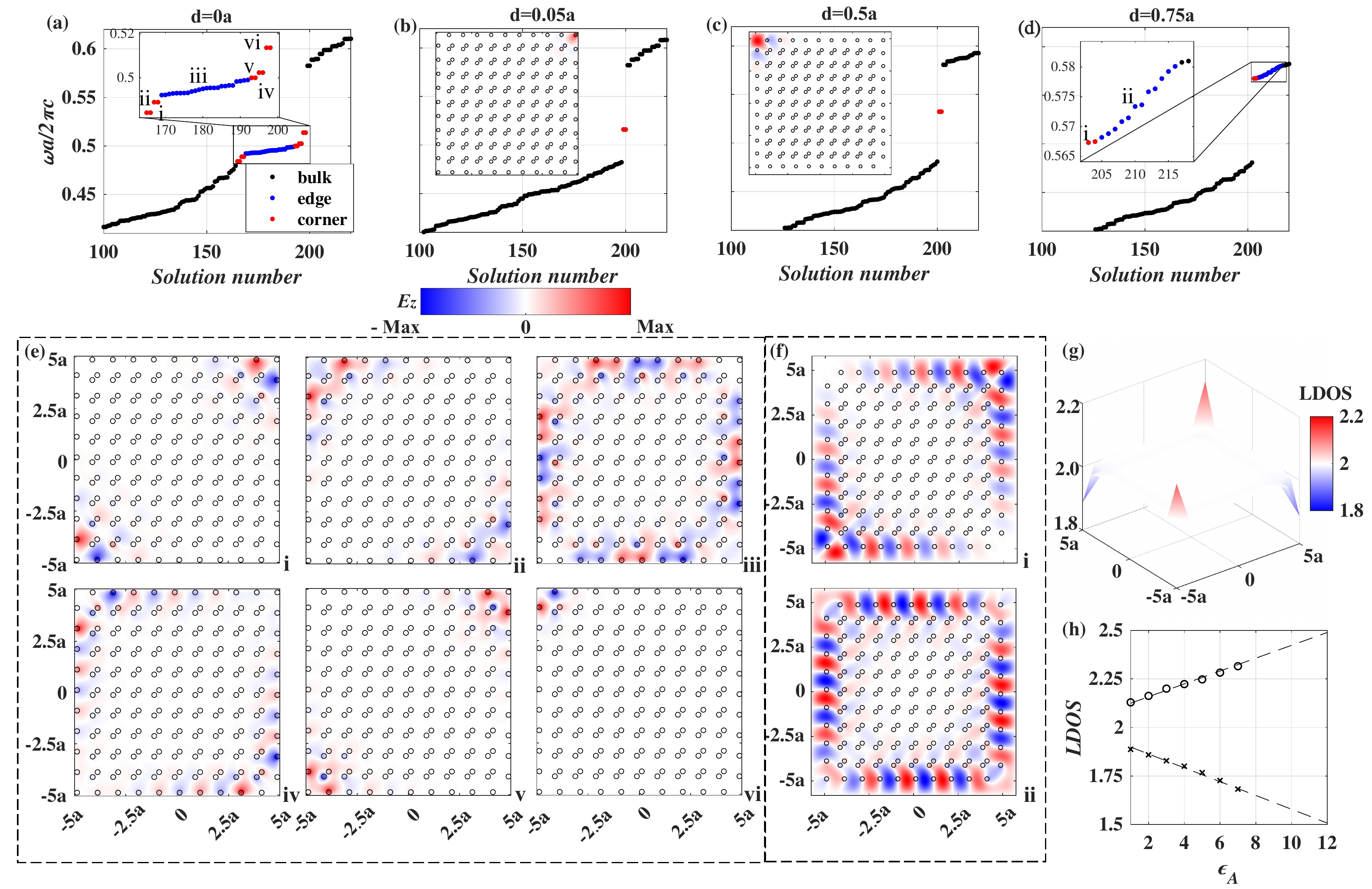}
	\caption{(a-d) Band spectrum of a supercell with $\varepsilon_A=1$, $\varepsilon_B=12$. PEC boundaries are used, and the distance between PEC and PhC are $d=0a$(a), $d=0.05a$(b), $d=0.5a$(c), and $d=0.75a$(d), respectively. The inserted figures of (b) and (d) show the spatial distributions of corner state. (e) The spatial distributions of corner and edge states which are marked in (a) with i-vi; (f) The spatial distributions of corner and edge states which are marked in (d) with i-ii; (g) The spatial distributions of LDOS with $d=0.5a$; (f) The corner charges versus the different $\varepsilon_A$ from $1$ to $7$, and the dashed line is the extrapolation from the corner charges.}\label{CornerStatePEC}
\end{figure*}

At the second step, we construct the model with the perfect electrical conductor (PEC) as the boundary of PhCs, shown in Fig. \ref{CornerStatePEC} and the results of local density of states (LDOS) \cite{liu2021bulk,xie2021higher} are used to strictly verify the topological origins of the two types of corner states. In Fig. \ref{CornerStatePEC}, we calculate the eigenfrequencies and eigenstates of a supercell which contains $N \times N$ unit cells with $\varepsilon_A=1$ and $\varepsilon_B=12$, and the distance between the PhC and PEC boundary is $d$. We choose $N=10$, and the band structures with $d=0a$, $0.05a$, $0.5a$, and $0.75a$ are shown in Fig. \ref{CornerStatePEC}(a-d), respectively. The typical localized states of Fig. \ref{CornerStatePEC}(a) and (d) are shown in Fig. \ref{CornerStatePEC}(e) and (f), respectively, which also sorted by the frequency from low to high. We will analyze a PhC with PEC boundary in the order of edge states, type-\uppercase\expandafter{\romannumeral2} corner states, and type-\uppercase\expandafter{\romannumeral1} corner states.

First, very interestingly, two sets of edge states would not appear at the same time when the PEC boundary is used. Specifically, DESs only appear with $d=0a$ as Fig. \ref{CornerStatePEC}(e)-iii shows, and AESs only appear with $d=0.75a$ as Fig. \ref{CornerStatePEC}(f)-ii shows.
The coupling between AESs or DESs will generate type \uppercase\expandafter{\romannumeral2} corner states, including all corner states in Fig. \ref{CornerStatePEC}(e) and (f).

Besides, some novelty phenomena about the type-\uppercase\expandafter{\romannumeral2} corner states in a square photonic lattice without both $C_4$ and $M_{x(y)}$ symmetries are observed for the first time.
For example, type-\uppercase\expandafter{\romannumeral2} corner states usually appear in pairs above and below the edge states, and they have opposite symmetry, such as state-i vs. state-v and state-ii vs. state-iv, as shown in Fig. \ref{CornerStatePEC}(e). Nevertheless, the corner states Fig. \ref{CornerStatePEC}(e)-vi and Fig. \ref{CornerStatePEC}(f)-i appear not in pairs. We believe that they should also have counterparts from the symmetry argument, but their counterparts are falling in the range of bulk states \cite{SI}.
Even more, from the field distribution shown in corner state ii and iv in Fig. \ref{CornerStatePEC}(e), we find that such type-\uppercase\expandafter{\romannumeral2} corner states are not very localized and physically they are from the larger long-range interactions between the rods in one cell. Such type-\uppercase\expandafter{\romannumeral2} corner states have not been found previously in kagome lattice \cite{li2020higher} and can also be observed by the tight-binding model if we introduce the long-range interactions into our tight-binding model (see supplementary S5 for details).

Second, we will investigate the corner states shown in Fig. \ref{CornerStatePEC}(b) and (c) with $d=0.05a$ and $d=0.5a$, which will be proved to be the topological non-trivial type-\uppercase\expandafter{\romannumeral1} corner states. Counterintuitively, except the corner states, there are no edge states inside the gap as shown in Fig. \ref{CornerStatePEC}(b) and (c). We can use the theory of surface impedance\cite{huang2016geometric, xiao2014surface,xiong2021resonance, li2019two, zhang2021fractal} to explain the absence of edge states (see supplementary information S5 for details). Two degenerated corner states in Fig. \ref{CornerStatePEC}(b) are concentrated on the dielectric rod, while the other two in Fig. \ref{CornerStatePEC}(c) are concentrated on the air. This phenomenon that only corner states exist in the whole gap range is reported for the first time to the best of our knowledge. Next, We will demonstrate the topological origin of such corner states in detail.

From the previous works of QTIs\cite{he2020quadrupole,benalcazar2019quantization}, there are several methods to judge the topological non-triviality of corner states. The first method is the filling anomaly that the non-trivial corner states are ``contributed" by both the top and bottom bands, respectively. In particular, a QTI with $N \times N$ cells, the solution numbers of non-trivial corner states are from $2N^2-1$ to $2N^2+2$ for the second gap. The second is the non-zero fractional corner charges from the corner states, which usually equal $0.5$ due to certain crystalline symmetries. As for the corner states of our system in Fig. \ref{CornerStatePEC}(b) and (c), we have counted the solution numbers of these corner states and find that they satisfy the filling anomaly. What's more, we calculate the spatial distribution for the sum of the lowest $2N^2$ energy states to obtain LDOS with $d=0.5a$ in Fig. \ref{CornerStatePEC}(g), and the non-zero fractional corner charges $Q_c$ could be found at the four corners, while the edge charges keep zero as same as traditional QTIs with $C_2$ symmetry (similar results also can be obtained with $d=0.05a$). Hence, the two judgements ensure that the corner states in \ref{CornerStatePEC}(b) and (c) are topologically non-trivial type-\uppercase\expandafter{\romannumeral1} corner states.

However, in contrast to traditional QTIs with $C_4$ and $M_{x(y)}$ symmetries, the fractional corner charges of our system are not equal to 0.5. This abnormal phenomenon can be explained by the absence of crystalline symmetries, e.g. the $C_4$ and $M_{x(y)}$ symmetries for our square lattice. We can confirm it by calculating corner charges versus the different $\varepsilon_A$ from $1$ to $7$, and the results are shown in Fig. \ref{CornerStatePEC}(h). The dashed line in Fig. \ref{CornerStatePEC}(h) is the extrapolation from our results. The extrapolation shows that the fractional corner charge would increase when we decrease the difference between $\varepsilon_A$ and $\varepsilon_B$, and $Q_c=0.5$ in a system with "perfect lattice symmetry" when $\varepsilon_A = \varepsilon_B$. It should be noted that actually we can NOT observe type-\uppercase\expandafter{\romannumeral1} corner states because of the degeneracies of the 2nd and 3rd bands for the systems with the perfect square lattice symmetry. So, such ideal $0.5$ corner charges only exist theoretically on the extrapolation line. Surprisingly, our symmetries-broken model provides a new method to realize type-\uppercase\expandafter{\romannumeral1} corner states despite quadrupole moment less than $0.5$ and we will show that such type-\uppercase\expandafter{\romannumeral1} corner states also have special properties compared with the common type-\uppercase\expandafter{\romannumeral1} corner states.

Different from common cases, from Fig. \ref{CornerStatePEC}(b) and (c) the first special property is that there are no edge states inside the gap and the topologically nontrivial type-\uppercase\expandafter{\romannumeral1} corner states are near the gap center which far away from the bands. This means that type-\uppercase\expandafter{\romannumeral1} corner states of our system are very localized which is proved by the field distributions in Fig. \ref{CornerStatePEC}(b) and (c). Hence, the type-\uppercase\expandafter{\romannumeral1} corner states of such systems could be widely used as high-Q cavities with sub-wavelength scale. What's more, the corner states in Fig. \ref{CornerStatePEC} (b) and (c) concentrate on dielectric and air regions, respectively. This property is also a fantastic result from the symmetry breaking. Such air-concentrated type-\uppercase\expandafter{\romannumeral1} corner state shown in Fig. \ref{CornerStatePEC} (c) is observed for the first time and it has advantages in the design of the cavities with very strong field and the detectors for liquid, molecules in the air, etc.

At last, we would note that those type-\uppercase\expandafter{\romannumeral2} corner states and non-trivial type-\uppercase\expandafter{\romannumeral1} corner states also could be realized by the simple tight binding model (TBM) without $C_4$ and $M_{x(y)}$ symmetries, which confirmed the universality of these corner states. The detailed derivations and results of TBM are given in supplementary information S5.

\section{conclusion}

In summary, we have constructed general valley TI and HOTI in PhCs without both $C_4$ and $M_{x(y)}$ symmetries but with time-reversal symmetry, and the physical origins of edge states and two types of corner states are demonstrated in detail.
Our results reveal rich topological physics beyond the ``perfect lattice symmetries", which contribute to our understanding of new topological physics and extend the methods to realize topological non-trivial phase and states. The proposed states of such all-dielectric PhCs, with excellent properties of low group velocity, sub-wavelength localization, and air-concentrated field distribution, can be widely realized at almost all frequencies of interest and be further utilized to design the aimed optoelectronic devices with enhanced robustness, such as topological slow light, Fano resonator, detector, switch, and laser, etc. Furthermore, the related topics in high-dimensional systems and other waves, e.g., phonons and electrons, are also very attractive.

\begin{acknowledgments}
	This work is supported by National High Technology Research and Development Program of China (17-H863-04-ZT-001-035-01); National Key Research and Development Program of China (2016YFA0301103, 2018YFA0306201). We thank Professor Wei E.I. Sha and Dr. Samuel J Palmer for their open source codes \cite{zhao2020first, palmer2020peacock}.
\end{acknowledgments}
	
\nocite{*}
\bibliographystyle{apsrev4-2}
\input{manuscript.bbl}

\end{document}

%% file: manuscript.bbl
%